\documentstyle[12pt]{article}

\begin{document}
\setcounter{page}{1} \pagestyle{plain} \vspace{1cm}
\begin{center}
\Large{\bf Statistical mechanics of ideal gas in the presence of minimal length and maximal momentum}\\
\small \vspace{1cm}
{ H. Shababi\footnote{hedie.shababi@gmail.com}}\\
\vspace{0.5cm} {\it Department of Physics, Islamic Azad
University,\\
Sari Branch, Sari, Iran}
\end{center}
\vspace{1.5cm}
\begin{abstract}
various  approaches to quantum gravity suggest that the fundamental
volume of the phase space of the given space for representative
points, means $ \omega_{0}$, should be modified. In this paper, we
study the effects of this modification on the thermodynamics of an
ideal gas within the micro canonical ensemble. For this end, we use
a Generalized Uncertainty Principle (GUP) that admits both a minimal
measurable length and a maximal momentum. Using this GUP causes
decreasing the total number of the microstates of the system. In the
first step, we calculate these reductions for classical ideal gas,
and
in the second step, we calculate these effects for ultra relativistic gas.
\end{abstract}
\vspace{1.5cm}
\newpage
\section{Introduction}
Different approaches to quantum gravity such as String Theory[1-5],
non commutative geometry[6], loop quantum gravity[7],  and Doubly
Special Relativity predict the existence of a minimal measurable
length or a maximal observable momentum[8,9]. These theories argue
that near the Planck scale, the Heisenberg Uncertainty Principle
should be replaced by the so called Generalized Uncertainty
principle(GUP)[10,11]. In these paper we want to study the effects
of this GUP on the statistical mechanics of ideal gas. In ordinary
standard statistical mechanics, it is impossible to define the
position of a representative point in the phase space of the given
system more accuracy than the situation which is given by $(\Delta x
\Delta p)\geq\hbar $. It means that around any point (q,p) of the
(two dimensional) phase space, there exist an area of the order of
$\hbar$ which can't say where the exact position of the particle is.
In $2N$ dimensional phase space, the corresponding volume of
uncertainty around any point would be of order $\hbar^{N}$. So we
can assume approximately that the phase space is made of elementary
cells of volume $\hbar^{N}$. These cells have one to one
correspondence with the quantum mechanics states of the given
system[12]. Not that an elegant formulation of statistical mechanics
of multi-dimensional Cantor sets based on fractal nature of
space-time has been provided by El Naschie[13]. Although, this issue
has been studied by some authors [14], but they ignored this fact
that a minimal measurable length essentially requires the existence
of a maximal momentum encoded in the duality of position-momentum
spaces or uncertainty principle. Existence of a maximal measurable
momentum for a test particle modifies the results of the mentioned
studies considerably. In which follow we consider the mentioned GUP,
as our primary input. We consider a micro-cononical ensemble and
then study the thermodynamical properties of an ideal gas composed
of monatomic non-interacting particles within  mentioned GUP. For
this reason, we study the total number of microstates of the given
system and find the novel and unusual results. One may ask about the
possible detection of these extra ordinary effects. Up to now, there
is no direct experimental or observational scheme for detecting of
these novel effects. Nevertheless, since the basis of our
calculations come back to the GUP, possible experimental schemes for
gravity predictions have been proposed[15]. Therefore any search for
quantum gravity signals provides possible indirect test of
generalized statistical mechanics which we want to consider. It
should be mentioned that these results are only important in the
limit of high momentum and without these consideration, we have the
results of normal statistical mechanics.

\newpage

\section{Thermodynamics of classical ideal gas in the presence of GUP}
In this section we use an ideal gas that composed of  monatomic
non-interacting particles within GUP that admits both minimal length
and maximal momentum. We suppose this ideal gas with the mentioned
situations, in a micro-canonic ensemble. In micro-canonical
ensemble, the macro state of the system is given by the number of
molecules $N$, the physical volume $V$, and energy $E$ of the
system. In this ensemble, the volume $\omega$ of the phase space to
the representative points of the system where lie anywhere within a
hypershell defined by the boundary condition $E-\frac{\Delta}{2}\leq
H(q,p)\leq E+\frac{\Delta}{2}$ is given by
\begin{equation}
\omega=\int' d\omega=\int'\int' (d^{3N}q)(d^{3N}p)
\end{equation}
where $\omega\equiv\omega(N,V,E;\Delta)$. In this equation the
primed symbol means that integral could extends only over that part
of the phase space which agrees to the above condition. Since the
Hamiltonian of the classical ideal gas is only function of the $ps$,
the integration over the $qs$ can be written as $V^{N}$. The
remaining integral
\begin{equation}
\int' (d^{3N}p)
\end{equation}
should be calculated under the following condition:
\begin{equation}
2m[E-\frac{\Delta}{2}]\leq \Sigma_{k=1}^{3N}p_{i}^{2}\leq
2m[E+\frac{\Delta}{2}]
\end{equation}
This condition is suitable only in standard framework. In this
paper, with considering a GUP in the presence of both minimal
measurable length and maximal momentum, the hypershell situation
should be modified. This GUP can be written as follows
\begin{equation}\Delta x\Delta
p\geq\frac{\hbar}{2}\bigg[1+\Big(\frac{\beta}{\sqrt{\langle
p^2\rangle}}+4\beta^{2}\Big)(\Delta p)^2+4\beta^{2}\langle
p\rangle^2-2\beta\sqrt{\langle p^2\rangle}\bigg].
\end{equation}
Since $(\Delta p)^2=\langle p^2\rangle-\langle p\rangle^2$,  by
setting $\langle p\rangle=0$ for simplicity, we find
\begin{equation}
\Delta x\Delta p\geq \frac{\hbar}{2}\bigg(1-\beta(\Delta
p)+2\beta^{2}(\Delta p)^2\bigg).
\end{equation}
which $\beta$ is a positive quantity.\\
 It is easy to show how this
setup leads to a maximal momentum. To show this end, we note that
the absolute minimal measurable length in our setup is given by,
$\Delta x_{min}(\langle p\rangle=0)=\frac{2\sqrt{2}-1)}{2}\hbar\beta
$. Due to duality of position and momentum operators, it is
reasonable to assume $\Delta x_{min}\propto\Delta p_{max}$.  Now,
saturating the inequality in relation (5), we find
\begin{equation}
2(\Delta x\Delta p)=\hbar\bigg(1-\beta(\Delta p)+2\beta^{2}(\Delta
p)^2\bigg).
\end{equation}
This results in
\begin{equation}
(\Delta p)^2-\frac{(2\Delta x+\beta\hbar)}{2\hbar\beta^{2}}(\Delta
p)+\frac{1}{2\beta^{2}}=0.
\end{equation}
So, we find
\begin{equation}
(\Delta p_{max})^2-\frac{(2\Delta
x_{min}+\beta\hbar)}{2\hbar\beta}(\Delta
p_{max})+\frac{1}{2\beta^{2}}=0.
\end{equation}
Now using the value of $\Delta x_{min}$, we find
\begin{equation}
(\Delta p_{max})^2-\frac{\sqrt{2}}{\beta}(\Delta
p_{max})+\frac{1}{2\beta^{2}}=0.
\end{equation}
The solution of this equation is
\begin{equation}
\Delta p_{max} =\frac{\sqrt{2}}{2\beta }.
\end{equation}
So, there is an upper bound on particle's momentum uncertainty. As a
nontrivial assumption, we assume that this maximal uncertainty in
particle's momentum is indeed the maximal measurable momentum. This
is of the order of Planck momentum. We note that neglecting a factor
of $\frac{1}{2}$ for simplicity in our forthcoming arguments, the
GUP formulated as (5) gives the following generalized commutation
relation
\begin{equation}
[x,p] = i\hbar\Big(1 - \beta p + 2\beta^{2} p^2\Big).
\end{equation}
With comparing this equation with its standard form we can interpret
this as a generalization of $\hbar$. This generalization has
important results. One of them is increasing the volume of phase
space, it means that the standard volume of phase space $\hbar^{N}$
changes to $[\hbar(1-\beta p+2\beta^{2}p^{2})]^{N}$. It is clear
that the nature result of increasing the volume of the phase space
is decreasing of the number of accessible microstates for the given
system. Now with using this GUP we should modified the hypershell
boundary condition. This issue firstly has been considered by
Kalyana Rama. He has discussed the effect of GUP on various
thermodynamical quantities in grand canonical ensemble[16]. Then
some authors considered the effects of GUP on thermodynamics of
ideal gas in micro-canonical ensemble[17]. But they didn't consider
the effects of maximal momentum in their calculations. In this
paper, we want to study this results  in the presence of  GUP that
formulated as (5). Within this GUP framework, particle's momentum
should be generalized. This generalized momentum is given by
\begin{equation}
p^{GUP}\simeq p(1-\frac{1}{2}\beta p +\frac{2}{3}\beta^{2}p^{2}).
\end{equation}
On the other hand, when the momentum be generalized, energy will
generalize too
\begin{equation}
E^{GUP}\simeq E(1-\frac{1}{2}\beta E +\frac{2}{3}\beta^{2}E^{2}).
\end{equation}
Now with using equations (12) and (13) we should rewrite the
standard hypershell equation. So we have
\begin{equation}
2m\bigg[E(1-\frac{1}{2}\beta E+\frac{2}{3}\beta^{2}
E^{2})-\frac{\Delta}{2}\bigg]\leq\sum_{i=1}^{3N}p_{i}^{2}(1-\beta
p_{i}+\frac{4}{3}\beta^{2} p_{i}^2)\leq 2m\bigg[E(1-\frac{1}{2}\beta
E+\frac{2}{3}\beta^{2} E^{2})+\frac{\Delta}{2}\bigg]
\end{equation}
In this situations, integral(2) is equal to the volume of a
3N-dimensional hypershell, bounded by two hyperspheres of radii
$$ \sqrt{2m\bigg[E(1-\frac{1}{2}\beta E+\frac{2}{3}\beta^{2}
E^{2})-\frac{\Delta}{2}\bigg]}$$ and
$$\sqrt{2m\bigg[E(1-\frac{1}{2}\beta E+\frac{2}{3}\beta^{2}
E^{2})+\frac{\Delta}{2}\bigg]}.$$ So we can write
\begin{equation}
\int'...\int'\prod_{i=1}^{3N}dp_{i}=B_{3N}\Bigg(\sqrt{2m[E^{GUP}+\frac{\Delta}{2}]}
\Bigg)^{3N}:=K
\end{equation}
This equation has written from the following condition
\begin{equation}
0\leq\sum_{i=1}^{3N} p_{i}^2(1-\beta
p_{i}+\frac{4}{3}\beta^{2}p_{i}^2)\leq 2m[E^{GUP}+\frac{\Delta}{2}]
\end{equation}
This equation gives half of the volume of the phase space, so we
should multiply our final result to the factor of $2$. Here $B_{3N}$
is a constant which depends only on the dimensionality of the given
phase space. Now, we can calculate the volume element $dK$ as
follows
\begin{equation}
dK=\frac{3}{2}N
B_{3N}(\sqrt{2m})^{3N}\Bigg[\sqrt{E^{GUP}+\frac{\Delta}{2}}\Bigg]^{3N-2}
dE^{GUP}
\end{equation}
To evaluate $B_{3N}$, we use the following integral formula
\begin{equation}
\int_{-p_{p}}^{+p_{p}} \exp(-p^{2}+\beta
p^{3}-\frac{4}{3}\beta^{2}p^{4})dp=R(\beta)
\end{equation}
As we see, when we use GUP that admits both minimal measurable
length and maximal momentum, the boundary condition  of the
integral, should be restricted. In this equation $p_{p}$ is the
Planck momentum and $R(\beta)$ is the result of integral. Because of
the complication of the solution of integral, first we expand the
exponential function up to eighth order and then calculate the
integral. This calculation give us
$$
R(\beta)=2p_{p}-\frac{2}{3}p_{p}^{3}+\frac{1}{2}\beta
p_{p}^{4}+\frac{2}{5}\bigg(\frac{1}{2}-\frac{4}{3}\beta^{2}\bigg)p_{p}^{5}-\frac{1}{3}\beta
p_{p}^{6}$$
\begin{equation}
+\frac{2}{7}\bigg(\frac{11}{6}\beta^{2}-\frac{1}{6}\bigg)p_{p}^{7}+\frac{1}{4}\bigg(\frac{1}{2}\beta-\frac{4}{3}\beta^{3}\bigg)p_{p}^{8}+...
\end{equation}
For simplicity, we supposed this calculation equal to $R(\beta)$.
Another way to calculate the above integral is considering only
first order of $\beta$, so we have
$$R(\beta)=\beta-\beta p_{p}^{2}e^{-p_{p}^{2}}+\sqrt{\pi}
erf(p_{p})$$ where $erf(x)$ is the error function and defined as
$$erf(x)=\frac{2(\int_{0}^{x}e^{-t^{2}}dt)}{\sqrt{\pi}}$$
 With multiplying $3N$ to such integrals, one
for each of variables $p_{i}$, we have
\begin{equation}
\int_{-p_{p}}^{+p_{p}}...\int_{-p_{p}}^{+p_{p}}\exp\Bigg(-\sum_{i=1}^{3N}p_{i}^{2}(1-\beta
p_{i}+\frac{4}{3}\beta^{2}p_{i}^2)\Bigg)\prod_{i=1}^{3N}
dp_{i}=[R(\beta)]^{3N}
\end{equation}
Therefore, it follows that:
\begin{equation}
\int_{-p_{p}}^{+p_{p}}\exp\Bigg(-2m(E^{GUP}+\frac{\Delta}{2})\bigg)dK=[R(\beta)]^{3N}
\end{equation}
Now we put Eq.(17) to the above equation and find
\begin{equation}
\int_{-p_{p}}^{+p_{p}}\frac{3}{2}N B_{3N}
(\sqrt{2m})^{3N}\bigg[\sqrt{E^{GUP}+\frac{\Delta}{2}}\bigg]^{3N-2}\exp\bigg(-2m(E^{GUP}+\frac{\Delta}{2})\bigg)dE^{GUP}=[R(\beta)]^{3N}
\end{equation}
In the next step, with calculating this integral, we can find
$B_{3N}$. So we have:
\begin{equation}
B_{3N}=\frac{2[R(\beta)]^{3N}\exp(m\Delta)}{3N(2m)^{\frac{3N}{2}}\int_{-p_{p}}^{+p_{p}}(E^{GUP}+\frac{\Delta}{2})^{\frac{3N-2}{2}}\exp(-2mE^{GUP})dE^{GUP}}
\end{equation}
For $\Delta\ll E^{GUP}$, this equation reduces to:
\begin{equation}
B_{3N}=\frac{2[R(\beta)]^{3N}}{3N(\frac{3N}{2}-1)!}
\end{equation}
Now, from equation (15), we find
\begin{equation}
\int...\int\prod_{i=1}^{3N}dp_{i}\equiv\frac{2[R(\beta)]^{3N}(2mE^{GUP})^{\frac{3N}{2}}\bigg[1+\frac{3N\Delta}{4E^{GUP}}\bigg]}{3N(\frac{3N}{2}-1)!}
\end{equation}
As we now ,for thermodynamical systems, $N\gg1$. So we can  rewrite
the above equation as follow
\begin{equation}
\int...\int\prod_{i=1}^{3N}dp_{i}\simeq
\frac{\Delta}{2E^{GUP}}\frac{[R(\beta)]^{3N}}{(\frac{3N}{2}-1)!}\bigg(2mE^{GUP}\bigg)^{\frac{3N}{2}}
\end{equation}
So, with these situations, the total volume of the phase space
enclosed within hypershell is given by
\begin{equation}
w\simeq\frac{\Delta}{E^{GUP}}V^{N}\frac{(2[R(\beta)]^{2}m
E^{GUP})^{\frac{3N}{2}}}{(\frac{3N}{2}-1)!}
\end{equation}
Now we want to find the number of microstates. To find this end,
first we should find the fundamental volume $\omega_{0}$ in the
presence of the minimal length and maximal momentum. It is given by
\begin{equation}
\omega_{0}=(\Delta q \Delta p)^{3N}=\bigg[\hbar(1-\beta
p+2\beta^{2}p^{2})\bigg]^{3N}\equiv\hbar'^{3N}
\end{equation}
In this equation, for simplicity we supposed
that$\bigg[\hbar(1-\beta p+2\beta^{2}p^{2})\bigg]\equiv \hbar'$.
With compering Eq.(28) with its standard form, we can consider it as
generalized of $ \hbar$. In other words, $\hbar\longrightarrow
\hbar(1-\beta p+2\beta^{2}p^{2})$. It is obvious that, with dividing
total volume phase space to fundamental phase space, means
$\frac{\omega}{\omega_{0}}$,  we can find total number of
microstates within hypershell and it has shown  by $\Omega$.\\
 So we have
\begin{equation}
\Omega=\frac{V^{N}}{\hbar'^{3N}}\frac{\Delta}{E^{GUP}}\frac{(2[R(\beta)]^{2}m
E^{GUP})^{\frac{3N}{2}}}{(\frac{3N}{2}-1)!}
\end{equation}
Obviously, within GUP in the presence of minimal length and maximal
momentum, due to increased fundamental volume $\omega_{0}$, the
number of total microstates decreases. \\
The complete thermodynamics of the given system would then given by,
\begin{equation}
S(N,V,E^{GUP})=k\ln\Omega=k\ln\Big(\frac{V^{N}}{\hbar'^{3N}}\frac{\Delta}{E^{GUP}}\frac{(2[R(\beta)]^{2}m
E^{GUP})^{\frac{3N}{2}}}{(\frac{3N}{2}-1)!}\Bigg)
\end{equation}
Where $S$ is the entropy of the given system. In the absence of
quantum gravity effects, when $\beta\longrightarrow0$, we have the
usual standard entropy. Now, after finding entropy, we can find
various thermodynamical quantities. it is obvious from above
equations that the reduction of  total number of accessible
microstates in high momentum regime cause reduction of entropy. It
seems that thermodynamical system in very short distances have an
unusual thermodynamics.
\section{Thermodynamics of extreme relativistic gas in the presence of GUP}
In this part, we want to calculate thermodynamics of an ultra
relativistic monatomic non interacting gaseous system in the
presence of GUP that admit both minimal measurable length and
maximal momentum. With using of arguments in previous  section, the
hypershell equation for ultra relativistic gaseous system is given
by the following equation
\begin{equation}
\frac{1}{c}\bigg[E\big(1-\frac{1}{2}\beta
E+\frac{2}{3}\beta^{2}E^{2}\big)-\frac{\Delta}{2}\bigg]\leq\sum_{i=1}^{3N}p_{i}\big(1-\frac{1}{2}\beta
p_{i}+\frac{2}{3}\beta^{2}p_{i}^{2}\big)\leq
\frac{1}{c}\bigg[E\big(1-\frac{1}{2}\beta
E+\frac{2}{3}\beta^{2}E^{2}\big)+\frac{\Delta}{2}\bigg]
\end{equation}
In this situation, $\int(d^{3N}p)$ is equal to the volume of a $3N$
dimensional hypershell, bounded by two hypershell of radii\\
$$\sqrt{\frac{1}{c}\bigg[E\big(1-\frac{1}{2}\beta
E+\frac{2}{3}\beta^{2}E^{2}\big)-\frac{\Delta}{2}\bigg]}$$
 and
$$\sqrt{\frac{1}{c}\bigg[E\big(1-\frac{1}{2}\beta
E+\frac{2}{3}\beta^{2}E^{2}\big)+\frac{\Delta}{2}\bigg]}$$ The
number of microstates for the system is proportional to the volume
of this hypershell. Similarly to the previous section, we have
\begin{equation}
\int'...\int'\prod_{i=1}^{3N}dp_{i}=A_{3N}\Big(\sqrt{\frac{1}{c}[E^{GUP}+\frac{\Delta}{2}]}
\Big)^{3N}:=F
\end{equation}
This equation follows the following condition
\begin{equation}
0\leq\sum_{i=1}^{3N} p_{i}\bigg(1-\frac{1}{2}\beta
p_{i}+\frac{2}{3}\beta^{2}p_{i}^{2}\bigg)\leq
\frac{1}{c}\bigg[E^{GUP}+\frac{\Delta}{2}\bigg]
\end{equation}
where $E^{GUP}$ is equal to $E\big(1-\frac{1}{2}\beta
E+\frac{2}{3}\beta^{2}E^{2}\big)$. This equation shows half of the
volume of the phase space. So we should multiply the final result
with a factor of $2$. Now we want to find $dF$:
\begin{equation}
dF=\frac{3}{2}N
A_{3N}(\frac{1}{c})^{3N}\bigg[\sqrt{E^{GUP}+\frac{\Delta}{2}}
\bigg]^{3N-2} dE^{GUP}
\end{equation}
To evaluate $A_{3N}$, we use the following integral formula:
\begin{equation}
\int_{-p_{p}}^{+p_{p}} \exp\bigg(-p+\frac{1}{2}\beta p^2
-\frac{2}{3}\beta^{2} p^{3}\bigg)dp=H(\beta)
\end{equation}
Here $H(\beta)$ is the result of the integral. In above equation
because of the complicated form of sentences, first we expand them
up to eighth order, and then calculate the integral, so we have:
$$H(\beta)=2p_{p}-p_{p}^{2}+\frac{2}{3}(\frac{1}{2}+\frac{1}{2}\beta)p_{p}^{3}+\frac{1}{2}(-\frac{2}{3}\beta^{2}-\frac{1}{2}\beta-\frac{1}{6})p_{p}^{4}$$
$$+\frac{2}{5}(\frac{19}{24}\beta^{2}+\frac{1}{4}\beta+\frac{1}{24})p_{p}^{5}+\frac{1}{3}(-\frac{1}{3}\beta^3-\frac{11}{24}\beta^2-\frac{1}{12}\beta-\frac{1}{120})p_{p}^{6}$$
$$+\frac{2}{7}(\frac{2}{9}\beta^{4}+\frac{17}{48}\beta^{3}+\frac{25}{144}\beta^{2}-\frac{1}{48}\beta+\frac{1}{720})p_{p}^{7}+\frac{1}{4}(-\frac{11}{36}\beta^{4}-\frac{3}{16}\beta^{3}$$
\begin{equation}
-\frac{7}{144}\beta^{2}-\frac{1}{240}\beta-\frac{1}{5040})p_{p}^{8}+...\quad\quad\quad\quad\quad\quad\quad\quad\quad\quad\quad\
\end{equation}
Similarly to the previous section, another way to calculate the
above integral is considering only first order of $\beta$. With this
situation, we have
$$H(\beta)=2+2\beta-2e^{-p_{p}}(1+\beta+\beta p_{p}+\frac{1}{2}\beta
p_{p}^{2})$$ Now with multiplying $3N$ to this integral, for each of
variable $p_{i}$, we obtain
\begin{equation}
\int_{-p_{p}}^{+p_{p}}...\int_{-p{p}}^{+p_{p}}\exp\Big(-\sum_{i=1}^{3N}p_{i}(1-\frac{1}{2}\beta
p_{i}+\frac{2}{3}\beta^{2}{p_{i}^{2}})\Big)\prod_{i=1}^{3N}dp_{i}=[H(\beta)]^{3N}
\end{equation}
Since, it follows that
\begin{equation}
\int_{-p_{p}}^{+p_{p}}
\exp\bigg(-\frac{1}{c}(E^{GUP}+\frac{\Delta}{2})\bigg)dF=[H(\beta)]^{3N}
\end{equation}
Now with putting Eq.(34) in the above equation, we obtain
\begin{equation}
\int_{-p_{p}}^{+p_{p}}\frac{3}{2}N
A_{3N}(\frac{1}{c})^{\frac{3N}{2}}\bigg[E^{GUP}+\frac{\Delta}{2}\bigg]^{\frac{3N-2}{2}}\exp\bigg(-\frac{1}{c}(E^{GUP}+\frac{\Delta}{2})\bigg)dE^{GUP}=[H(\beta)]^{3N}
\end{equation}
Therefore we can easily find $A_{3N}$
\begin{equation}
A_{3N}=\frac{2[H(\beta)]^{3N}\exp(\frac{\Delta}{2c})}{3N(\frac{1}{c})^{\frac{3N}{2}}\int_{-p_{p}}^{+p_{p}}\big[E^{GUP}+\frac{\Delta}{2}\big]^{\frac{3N-2}{2}}\exp\big(-\frac{1}{c}E^{GUP}\big)dE^{GUP}}
\end{equation}
For $\Delta\ll E^{GUP}$, this equation equals to
\begin{equation}
A_{3N}=\frac{2[H(\beta)]^{3N}}{3N (\frac{3N}{2}-1)!}
\end{equation}
Now from Eq.(32), we can obtain
\begin{equation}
\int...\int\prod_{i=1}^{3N}dp_{i}=\frac{2[H(\beta)]^{3N}(\frac{1}{c}E^{GUP})^{\frac{3N}{2}}\big[1+\frac{3N\Delta}{4E^{GUP}}\big]}{3N
(\frac{3N}{2}-1)!}
\end{equation}
On the other hand, when $N\gg1$, we can write
\begin{equation}
\int...\int\prod_{i=1}^{3N}dp_{i}\simeq
\frac{\Delta}{2E^{GUP}}\frac{[H(\beta)]^{3N}}{(\frac{3N}{2}-1)!}(\frac{1}{c}E^{GUP})^{\frac{3N}{2}}
\end{equation}
So the total volume of the phase space enclosed within hypershell is
given by
\begin{equation}
\omega\simeq\frac{\Delta}{E^{GUP}}V^{N}\frac{\big(\frac{1}{c}[H(\beta)]^{2}E^{GUP}\big)^{\frac{3N}{2}}}{(\frac{3N}{2}-1)!}
\end{equation}
As we know, the total number of microstates is given by $\Omega=
\frac{\omega}{\omega_{0}}$, and from Eq.(28), we have the
$\omega_{0}$ value in the GUP that admit both minimal length and
maximum momentum. So we have
\begin{equation}
\Omega=\frac{V^{N}}{\hbar'^{3N}}\frac{\Delta}{E^{GUP}}\frac{\big(\frac{1}{c}[H(\beta)]^{2}E^{GUP}\big)^{\frac{3N}{2}}}{(\frac{3N}{2}-1)!}
\end{equation}
Now, after finding total number of microstates, it is easy to find
thermodynamical quantities of the given system. In this step, the
first thermodynamical quantity we find is entropy. So
\begin{equation}
S(N,V,E^{GUP})=k\ln\Omega=k\ln\Big(\frac{V^{N}}{\hbar'^{3N}}\frac{\Delta}{E^{GUP}}\frac{\big(\frac{1}{c}[H(\beta)]^{2}E^{GUP}\big)^{\frac{3N}{2}}}{(\frac{3N}{2}-1)!}\Big)
\end{equation}
As we know, entropy is directly depend on the total accessible
number of phase space, so decreasing of the total number of
microstates cause reduction of entropy, and this is the unusual
behavior of thermodynamical quantities near the planck scale. In
standard situation, when $\beta\longrightarrow 0$, we have the
results of ordinary statistical mechanics. Various thermodynamical
quantities can then be calculated with using of Eq.(46).\\
\section{Summary and conclusion}
In this paper we consider a GUP that admits both minimal measurable
length and maximal momentum. In the presence of these GUP we studied
 some thermodynamical properties of an ideal gas both in the
classical and ultra relativistic limit. We have found that in the
presence of this mentioned GUP, the fundamental volume of the phase
space increased. We can interpret this increasing as a generalized
of the Planck's constant. This increasing of the volume caused
reduction of the accessible number of phase space and because of
this reduction, some important thermodynamical quantities such as
entropy of ideal gas decreased. It was an unusual behavior of
thermodynamics
 in very short distances.\\
 
{\bf Acknowledgment}\\
I would like to thank Prof. Kourosh Nozari for insightful comments
and discussion.


\begin{thebibliography}{12}
\bibitem{1}
 Veneziano G. A stringy nature needs just two constants.
Europhys Lett 1986;2:199;G. Veneziano, Proc Texas Superstring
Workshop, 1989
\bibitem{2}
 Amati D, Ciafaloni M, Veneziano G. Can spacetime be probed
below the string size? Phys Lett B 1989;216:41
\bibitem{3}
 Amati D, Ciafaloni M, Veneziano G. Superstring collisions at
Planckian energies. Phys Lett 1987;B197:81;Amati D, Ciafaloni M,
Veneziano G. Classical and quantum gravity effects from Planckian
energy superstring collisions. Int J Mod Phys A 1988;7:1615; Amati
D, Ciafaloni M, Veneziano G. Higher-order gravitational deflection
and soft bremsstrahlung in Planckian energy superstring collisions.
Nucl Phys B 1990;347:530
\bibitem{4}
 Gross DJ, Mende PF. String theory beyond the Planck scale.
Nucl Phys B 1988;303:407
\bibitem{5}
 Konishi K, Paffuti G, Provero P. Minimum physical length and
the generalized uncertainty principle in string theory. Phys Lett B
1990;234:276
\bibitem{6}
 Capozziello S, Lambiase G, Scarpetta G. Generalized
uncertainty principle from quantum geometry. Int J Theor Phys
2000;39:15
\bibitem{7}
 Garay LJ. Quantum gravity and minimum length. Int J Mod Phys A
1995;10:145
\bibitem{8}
J.Magueijo and L.Smolin
Phys.Rev.Lett.88(2002)190403,[arXiv:hep-th/0112090]; J.Magueijo and
L.Smolin Phys.Rev. D 71 (2005)026010 [arXiv:hep-th/0401087];
\bibitem{9}
J.L.Cortes,J.Gamboa,Phys.Rev.D 71 (2005) 065015
[arXiv:hep-th/0405285];
\bibitem{10}
{\raggedright Kempf.A, Mangano.G  and  Mann.R.B.: Hilbert space
representation of the minimal } {\raggedright length uncertainty
relation, Phys. Rev. D 52 (1995) 1108 }
\bibitem{11}
{\raggedright Nozari,K.,Pedram,P.: Minimal length and bouncing
particle spectrum, Europhys. } {\raggedright Lett. 92 (2010) 50013 }
\bibitem{12}
 Pathria RK. Statistical mechanics. 1st ed. Pergamon Press;
1972
\bibitem{13}
El Naschie MS. Statistical mechanics of multi-dimensional cantor
sets, Godel theorem and quantum spacetime. J Franklin Inst
1993;330:199-211
\bibitem{14}
Nozari,K,Mehdipour,S. Hamid. Chaos, Solitons and Fractals 32 (2007)
1637-1644
\bibitem{15}
 Amelino-Camelia G et al.Phys Rev D 2004;70:107501
 \bibitem{16}
Kalyana Rama S. Some consequences of the generalized uncertainty
principle: statistical mechanical, cosmological, andvarying speed of
light. Phys Lett B 2001;519:103-10
\bibitem{17}
 Nozari,K, Mehdipour,S. Hamid. Chaos, Solitons and Fractals 32
(2007) 1637-1644

\end{thebibliography}
\end{document}